\documentstyle[aps]{revtex}
\include{epsf}

\begin{document}

\draft

\title{Critical gravitational collapse with angular momentum: from
critical exponents to scaling functions}

\author{Carsten Gundlach}

\address{Enrico Fermi Institute, University of Chicago, 5640 S. Ellis
Avenue, Chicago, IL 60637} 
\address{ Faculty of Mathematical Studies,
University of Southampton, Southampton SO17 1BJ, United
Kingdom\footnote{Current address}}

\date{9 October 2001}

\maketitle


\begin{abstract}

We investigate the threshold of gravitational collapse with angular
momentum, under the assumption that the critical solution is spherical
and self-similar and has two growing modes, namely one spherical mode
and one axial dipole mode (threefold degenerate). This assumption
holds for perfect fluid matter with the equation of state
$p=\kappa\rho$ if the constant $\kappa$ is in the range
$0<\kappa<1/9$. There is a region in the space of initial data where
the mass and angular momentum of the black hole created in the
collapse are given in terms of the initial data by two universal
critical exponents and two universal functions of one argument. These
expressions are similar to those for the correlation length and the
magnetization in a ferromagnet near its critical point, as a function
of the temperature and the external magnetic field. We discuss
qualitative features of the scaling functions, and hence of critical
collapse with high angular momentum.

\end{abstract}




\tableofcontents


\section{Introduction}
\label{section:introduction}


An isolated system in general relativity ends up in one of three final
states. It either collapses to a black hole, forms a star, or
disperses completely. The phase space of isolated gravitating systems
is therefore divided into basins of attraction. One cannot usually
tell into which basin of attraction a given data set belongs by any
other method than evolving it in time to see what its final state
is. The study of these boundaries in phase space, in particular of the
boundary between black hole formation and dispersion, is the subject
of the new field of critical collapse
\cite{Choptuik_review,critreview2}.

The pioneering work of Choptuik \cite{Choptuik} has shown that the
black hole threshold is both richer in structure and simpler than
naively expected. Choptuik carried out systematic high precision
numerical collapse simulations in the toy model of a spherically
symmetric massless scalar field coupled to general relativity. He
explored the black hole threshold by means of smooth one-parameter
families of initial data. A generic family contains both collapsing
and dispersing data sets. Choptuik found that locally each family
crosses the black hole threshold only once. This suggests that the
collapse threshold is a smooth hypersurface in phase space.  

The black hole mass shows a universal power-law scaling as a function
of distance from the black hole threshold, as measured by the
parameter $p$ of the one-parameter family of initial data. Let a black
hole be formed for $p>p_*$, and let dispersion occur for $p<p_*$. The
critical value $p_*$ of the parameter $p$ depends on the family, and
can only be determined through a bisection search. Choptuik found that
the black hole mass is approximately
\begin{equation}
\label{simplepower}
M(p) \simeq C (p-p_*)^\gamma \quad \text{for $p>p_*$},
\end{equation}
where the critical exponent $\gamma$ is independent of the family
(universal), with a numerical value of $\gamma\simeq 0.374$ for the
scalar field.

Furthermore, all evolutions with $p\simeq p_*$, on either side of
$p_*$, and for any family, pass through an intermediate attractor,
before finally dispersing or forming a black hole. This ``critical
solution'' is self-similar. In the case of the scalar field, the
self-similarity is a discrete symmetry. In the case of the perfect
fluid, it is continuous. By virtue of being an intermediate attractor,
the critical solution has precisely one growing mode. The critical
exponent for the black hole mass can be calculated from this growing
mode.

Subsequent work has found similar critical phenomena in many other
matter models (see \cite{critreview2} for a review). Unfortunately,
most work so far has been limited to a spherically symmetric,
uncharged situation in which a Schwarzschild black hole is
formed. Generic black holes, however, have angular momentum $L$ and
electric charge $Q$ as well as mass $M$, and we do not yet know what
happens if one fine-tunes to the black hole threshold along a
one-parameter family of data with significant angular momentum and/or
charge. 

Today, the role of angular momentum and charge in critical collapse is
understood only in the limit where they are so small that they can be
treated as linear perturbations of a spherically symmetric and
uncharged scenario throughout the collapse, from the initial data to
the final black hole. One can then make a connection from a small
perturbation in the initial data representing charge or angular
momentum to a linear perturbation of the final Schwarzschild black
hole that takes it into a Kerr-Newman black hole.

Specifically, critical collapse with a small amount of electric charge
has been investigated in the model of a spherically symmetric complex
scalar field coupled to a Maxwell field. A critical exponent for the
black hole charge $Q$ has been calculated by keeping track of the
least slowly decaying charged perturbation of the critical solution
\cite{chargepaper}. This prediction has subsequently been verified in
collapse simulations \cite{HodPirancharge}.

Similarly, critical collapse with a small amount of angular momentum
has been investigated in the model of a perfect fluid with equation of
state $p=\kappa\rho$, where $p$ is the pressure, $\rho$ the total
energy density, and $\kappa$ is a constant with $0<\kappa<1$. In exact
spherical symmetry, the critical exponent $\gamma$ for the black hole
mass is again independent of the initial data but depends on $\kappa$
\cite{EvansColeman,KoikeHaraAdachi2,Maison,NeilsenChoptuik}.
Perturbing around spherical symmetry, a critical exponent for black
hole angular momentum $L$ was calculated from the least slowly
decaying axial dipole perturbation in \cite{angmom}. The numerical
value of the critical exponent has been corrected in
\cite{critfluid2}. For the scalar field, a critical exponent for $L$
was calculated in second order perturbation theory around spherical
symmetry \cite{critscalangmom}. Both these predictions still await
testing, as high-precision simulations of rotating collapse to a black
hole are not yet available.

In this paper we go beyond the assumption that angular momentum is a
small perturbation throughout. This is necessary because the corrected
results for the axial dipole perturbations of the critical fluid in
\cite{critfluid2} show that all such perturbations decay only for the
equations of state $k>1/9$. For $k<1/9$, there is actually a single
growing mode. This means that in the limit in which initial data with
a small amount of angular momentum are fine-tuned to the black hole
threshold, there are two competing growing modes of the critical
solution, one spherical and one to do with differential rotation. The
latter is threefold degenerate because rotation is an axial
vector. Which of the two modes first reaches nonlinearity depends on
their initial amplitude and growth rate. In particular, if the initial
data are sufficiently close to the collapse threshold, rotation may
dominate the late stages of the evolution even if it was only a small
perturbation in the initial data, and the final black hole could be
very small but rapidly rotating.

We begin our presentation in section \ref{section:dynamicalsystem} by
reviewing the qualitative picture of critical collapse in terms of a
dynamical system, and establishing notation. In section
\ref{section:outline} we identify an intermediate attractor through
which solutions near the black hole threshold are funnelled. In
section \ref{section:initialdata} we discuss the initial data that
reach this attractor, and in \ref{section:scalingfunctions} we follow
them from the intermediate attractor to the black hole end state. We
find that the black hole mass and angular momentum depend on the
initial data only through universal functions of one combination of
the initial data parameters.  In this paper, we do not calculate these
``universal scaling functions'', but their mere existence
substantially constrains the phenomenology of critical collapse
resulting in small but rapidly spinning black holes.  We discuss this
phenomenology in section \ref{section:largeangmom}. The present work
has been motivated by the close analogy between critical collapse and
critical phase transitions in statistical mechanics, and represents an
extension of that analogy. This analogy is discussed separately in
section \ref{section:statmechanalogy}. We summarize in section
\ref{section:conclusions}.

In talking about black holes, it is customary to refer not to its
angular momentum $L$ but to its specific angular momentum $a\equiv
L/M$, and we adopt this convention here.


\section{The dynamical systems picture} 
\label{section:dynamicalsystem}


The time evolution of general relativity can be considered as an
(infinite-dimensional) dynamical system once the coordinate freedom of
general relativity has been fixed by imposing suitable coordinate
conditions on parts of the metric. Here we review this phase space
picture in preparation for the calculation of critical exponents and
scaling functions in the next section. Black holes on the one side,
and flat spacetime on the other, are attracting fixed points. Their
basins of attraction are separated by the black hole threshold, or
critical surface. It is obviously a hypersurface of codimension
one. The numerical evidence is consistent with the assumption that it
is smooth, and not, for example, fractal. By definition, the critical
surface is a closed dynamical system in its own right. Within the full
phase space, it is a repeller. Its attracting fixed points or limit
cycles are then attractors of codimension one in the full phase space,
with exactly one growing perturbation mode. They are called critical
fixed points in dynamical systems language, or critical solutions in
spacetime language.

Any trajectory beginning near the critical surface, but not
necessarily near the critical point, moves almost parallel to the
critical surface towards the critical point. As the critical point is
approached, the parallel movement slows down, and the phase point
spends some time near the critical point. Then the phase space point
moves away from the critical point in the direction of the growing
mode, and finally ends up on a stable fixed point. This is the origin
of universality: any initial data set that is close to the black hole
threshold (on either side) evolves to a spacetime that approximates
the critical spacetime for some time. When it finally approaches
either empty space or a black hole it does so on a trajectory that
appears to be coming from the critical point itself.

Not surprisingly, attractors of this dynamical system have higher
symmetry, when considered as Cauchy data for a spacetime, than generic
points in the phase space. The critical solutions found at the black
hole threshold, at least those known to date, are either static,
periodic, continuously self-similar, or discretely self-similar. Here
we consider only the continuously self-similar critical solutions,
which give rise to scaling laws of the form (\ref{simplepower}). The
model we focus on here, the perfect fluid with equation of state
$p=\kappa\rho$, has this type of critical solution. A
spacetime is continuously self-similar, or homothetic, if there exists
a vector field $\chi$ such that the Lie derivative of the spacetime
metric along $\chi$ obeys
\begin{equation}
{\cal L}_\chi g_{ab} = -2 g_{ab}.
\end{equation}
$\chi$ is called a homothetic vector field. It is a special case of
a conformal Killing vector field. By the Einstein equations,
continuous self-similarity implies that the matter stress-energy obeys
${\cal L}_\chi T_{ab} = 0$. 

In general relativity, each data set locally determines a unique
spacetime. But a given spacetime can be broken up into a sequence of
data sets, that is, foliated by spacelike hypersurfaces, in an
infinite number of ways, even if the initial hypersurface is kept
fixed. This means that the same physical spacetime can be described by
an infinite number of curves through the phase space of general
relativity. Turning general relativity into a dynamical system
therefore requires some prescription that returns a lapse and shift
for each point in phase space. In the following, let $\tau$ be the
time coordinate of the dynamical system, and let $x$ be the three
spatial coordinates. 

Calculations in critical phenomena require a prescription in which the
self-similar spacetimes are fixed points, and discretely self-similar
spacetimes are limit cycles. This means that for those sets of Cauchy
data that will evolve into a self-similar spacetime, the prescription
returns a lapse $\alpha$ and shift vector $\beta^a$ such that the time
vector $\partial/\partial \tau$ is the homothetic vector $\chi$:
\begin{equation}
\left({\partial\over\partial\tau}\right)^a\equiv\alpha n^a+\beta^a=\chi^a,
\end{equation}
where $n^a$ is the unit normal vector on the constant $\tau$
slices. Certain possible prescriptions, such as ``$K$-freezing lapse''
with ``minimal distortion shift'' have been identified \cite{GarGund},
and we now assume that one such choice has been made. Note that while
$\partial/\partial \tau$ becomes spacelike at large distances from the
center of collapse, this does not preclude surfaces of constant $\tau$
from being everywhere spacelike. The spacetime metric in these
coordinates is 
\begin{equation}
g_{\mu\nu}(\tau,x) = l^2 e^{-2\tau} \bar g_{\mu\nu}(x),
\end{equation}
where $\bar g_{\mu\nu}$ does not depend on $\tau$, and where $l$ is an
arbitrary fixed length scale.  Foliating by surfaces of constant
$\tau$ we obtain an induced metric and intrinsic curvature of the form
\begin{equation}
g_{ij}(\tau,x)=l^2 e^{-2\tau}\bar g_{ij}(x), \qquad
K_{ij}(\tau,x)=l e^{-\tau} \bar K_{ij}(x).
\end{equation}
We now turn this around to define the rescaled dynamical variables 
\begin{equation}
\bar g_{ij}(\tau,x)\equiv l^{-2} e^{2\tau}g_{ij}(\tau,x,), \qquad
\bar K_{ij}(\tau,x)\equiv l^{-1} e^{-\tau} K_{ij}(\tau,x)
\end{equation}
for the metric, and similar rescaled variables for the matter
fields. For a perfect fluid with 4-velocity $u^\mu$ and comoving
energy density $\rho$, for example, we define
\begin{equation}
\bar u^i\equiv l e^{-\tau}u^i, \qquad \bar\rho \equiv l^2 e^{-2\tau}\rho.
\end{equation}
In the variables $Z\equiv\{\bar g_{ij},\bar K_{ij},\bar
u^i,\bar\rho,\dots\}$, and with a suitable choice of lapse and shift,
a CSS solution will take the form $Z(x,\tau)=Z_*(x)$.  Full physical
initial data consist of $Z(x,\tau_0)$ and the value of $\tau_0$
itself, which provides an overall scale and allows one to reconstruct
the physical variables from the barred variables. 

It is helpful to think of the coordinates $x$ and $\tau$ and the
barred variables as dimensionless, of $le^{-\tau}$ as having a
dimension of length (or time, or mass, in units $c=G=1$), and of the
unbarred quantities as having their natural dimensions, with
$g_{\mu\nu}$ having a dimension of $l^2$ because the coordinates are
dimensionless. The Einstein and matter equations in the barred
variables are much like the usual ones, but are
dimensionless. $e^{-\tau}$ only appears in the combination
$le^{-\tau}$. In vacuum gravity, with a minimally coupled massless
scalar field, or with a perfect fluid with equation of state
$p=\kappa\rho$ fluid, no dimensionful constants appear in the field
equations. Therefore there are no constants to cancel $l$, and so no
powers of $e^{-\tau}$ can appear explicitly in the dimensionless
equations in the barred variables. This means that solutions are
possible which do not depend on $\tau$, and so are self-similar. If
dimensionful constants of dimension $l^{-n}$, with $n>0$, appear in
the field equations (for example, a mass term in the scalar wave
equation), they are multiplied by $l^ne^{-n\tau}$ in the
scale-invariant equations, and so become dynamically irrelevant as
$\tau\to\infty$, which corresponds to very small
scales. This allows for solutions that become asymptotically
self-similar on small scales (large $\tau$).


\section{The intermediate linear regime} 
\label{section:outline}


For what follows it is helpful to reformulate the mass-scaling law in
spherical symmetry, Eq. (\ref{simplepower}), by absorbing the
family-dependent constants $C$ and $p_*$ into a new parameter $\bar p$
that is a linear function of the parameter $p$, so that the law is now
\begin{equation}
\label{simplepower2}
M\simeq \bar p^\gamma \quad \text{for $\bar p>0$}
\end{equation}
for every 1-parameter family of initial data.  It is an experimental
fact that this is possible, with each data set assigned only one value
of $\bar p$, independently of the 1-parameter family of which it is
considered a part. This means that $\bar p$ is a scalar function on
the infinite-dimensional phase space, independently of any 1-parameter
families. $M$ is also a scalar, but it is not regular at the black
hole threshold, while $\bar p$ is regular with non-vanishing
gradient. $\bar p$ is therefore a good coordinate on phase space in a
neighborhood of the black hole threshold. This clarifies the invariant
meaning of the critical exponent $\gamma$: it is the unique power that
relates the observed scalar $M$ on phase space to another scalar $\bar
p$ that is a good coordinate.

Based on the phase space picture, the critical exponent $\gamma$ for
the black hole mass can be calculated in a manner suggested by Evans
and Coleman \cite{EvansColeman} and spelled out by Koike, Hara and
Adachi \cite{KoikeHaraAdachi1} and Maison \cite{Maison}. Here we
repeat this calculation, but now including the effects of angular
momentum in the initial data. We do this in the abstract notation $Z$
which applies not only to the perfect fluid with $k<1/9$, but equally
to any other model that has a spherically symmetric, continuously
self-similar solution with precisely two growing perturbation
modes. One of these must be spherically symmetric, and the other one
must have an axial dipole ($l=1$) angular dependence. All other
spherical and nonspherical perturbation modes (there are infinitely
many) must decay. The generalization from a continuously self-similar
to a discretely self-similar critical solution is trivial.

As the background solution we consider is continuously self-similar,
with $Z(x,\tau)=Z_*(x)$, the perturbations of this background depend
on $\tau$ only exponentially. We can write a generic perturbation on
the background $Z_*(x)$ as a sum over perturbation modes of the form
$e^{\lambda_i\tau}Z_i(x,\Omega)$, where the index $i$ labels the
perturbation spectrum. The two growing modes in particular are
designated $Z_0(x)$ (for the spherically symmetric one) and $\vec
Z_1(x,\Omega)$ (for the axial dipole mode). Here the arrow indicates
that $\vec Z_1$ is 3-fold degenerate, for $m=-1,0,1$. $\Omega$ is
shorthand for the angular dependence on the coordinates
$(\theta,\varphi)$. Perturbing around spherical symmetry, to linear
order rotation is associated with the axial dipole mode, and in
particular, the angular momentum of the spacetime is proportional to
the $r^{-2}$ falloff of the dipole mode at spacelike infinity. This is
true for regular perfect fluid spacetimes, but also for black holes:
outside the horizon, a Kerr black hole with $a\ll M$ can be written as
an axial dipole linear perturbation of a Schwarzschild black hole. The
significance of the axial dipole perturbation is discussed in more
detail in Refs.  \cite{critscalangmom} and \cite{fluidpert}. As a
matter of notation, we go from the basis $m=-1,0,1$ to the basis $xyz$
of $l=1$ harmonics, so that $\vec Z_1$ is an axial vector in 3-space
that points in the direction of angular velocity.

In this paper, we consider solutions that go through an intermediate
time regime in which they are well approximated by a self-similar,
spherically symmetric solution plus small perturbations, the
``intermediate linear regime''. Furthermore, in this regime, we
neglect the decaying linear perturbation modes, so that we have the
approximation
\begin{equation}
\label{intermediate_linear}
Z(x,\tau,\Omega) \simeq Z_*(x) + A e^{\lambda_0\tau}Z_0(x) + \vec B\cdot
e^{\lambda_1\tau} \vec Z_1(x,\Omega)+ \hbox{decaying modes}.
\end{equation} 
In the following we show that all solutions in a ball straddling the
hole threshold are in fact funneled through the intermediate linear
regime. Working backwards, in section \ref{section:initialdata} we
construct families of non-spherically symmetric data that go through
the intermediate linear regime. Working forwards, in section
\ref{section:scalingfunctions}, we extract Cauchy data in the
intermediate linear regime, and predict the mass and specific angular
momentum of the black hole from these intermediate data.


\section{Initial data with rotation}
\label{section:initialdata}


We now construct generic families of initial data that reach the
intermediate linear regime (\ref{intermediate_linear}). To do this, we
shall first introduce a parameter controlling angular momentum, and
then add a second, generic, parameter which allows us to tune to the
black hole threshold.

Consider first infinitesimally small deviations from a spherically
symmetric data set that does form a black hole. (For the moment we do
not assume that we are at the black hole threshold.) Perturbations
with different multipoles, and axial and polar perturbations, then
remain decoupled throughout the evolution, and so the angular momentum
of the final black hole is linearly related to the axial dipole part
of the initial data. However, while the specific angular momentum of
the black hole is a vector $\vec a$ (three numbers), the axial dipole
part of the initial data for a perfect fluid consists of a vector of
functions $\vec\beta(r)$, which characterize differential rotation. We
can explore this $3\infty$-dimensional space by means of many
different 3-parameter families of initial data, with the vector-valued
parameter $\vec q$ chosen so that $\vec q=0$ corresponds to spherical
symmetry, and so that $\vec q\to -\vec q$ corresponds to
$\vec\beta(r)\to -\vec\beta(r)$ for all $r$. Clearly, this is a
sufficient (but by no means a necessary) criterium for the relation
$\vec a(-\vec q)=- \vec a(\vec q)$ in the final black hole. It is also
clear that we have $M(-\vec q)=M(\vec q)$ for the black hole mass. (In
fact, from this it follows that to linear order in perturbation
theory, $M$ does not depend on $\vec q$.

But we need not stop at perturbations of spherical symmetry. In order
to make Schwarzschild black holes, one does not need spherically
symmetric initial data. Note that a Schwarzschild black hole has three
commuting reflection symmetries (also called octant symmetry), while a
Kerr black hole has only one reflection symmetry (through the
equatorial plane). Any reflection symmetry in the initial data is
maintained during evolution. Octant symmetry in the initial data is
therefore a sufficient (but not a necessary) criterium for producing a
non-rotating black hole. This leads us to consider 3-parameter
families of initial data that are not a perturbation of spherical
symmetry: we construct them so that the data with $\vec q=0$ are
octant-symmetric, and that data with $\vec q\ne 0$ contain a point
such that $\vec q\to -\vec q$ corresponds to a reflection through that
point. This symmetry between $\vec q$ and $-\vec q$ is conserved by
the time evolution, and so again we have the relations
\begin{equation}
\label{qdef}
\vec a(-\vec q)=-\vec a(\vec q), \qquad M(-\vec q)=M(\vec q).
\end{equation}
While octant symmetry of the initial data is a sufficient criterium
for the absence of angular momentum, it seems unlikely that one can
give a {\it necessary} criterium for the final black hole to have zero
angular momentum. But we can turn the argument around and formally
{\it define} the parameter $\vec q$ to have the properties
(\ref{qdef}). Such a definition may seem circular, but it is in fact
the analogue of the definition of the scalar function $\bar p$, in
that from the family-dependent parameter $\vec q$ we can define a
parameter $\bar{\vec q}$ that is a scalar on the space of initial data
and which is related in a simple universal way to the black hole
specific angular momentum $\vec a$.

We have defined the vector parameter $\vec q$ without assuming that we
were close to the black hole threshold. However, if the black hole
threshold is a surface of codimension one in the full space of
non-spherical initial data, then fine-tuning any other parameter $p$
in the initial data should take us to the black hole threshold, for
any fixed value of $\vec q$. With this motivation, consider now
4-parameter families of initial data with the following properties:

1. The initial data depend analytically on $p$ and $\vec q$ in a
   neighborhood of the black hole threshold. 

2. If for any $p$ and $\vec q$ a black hole is formed, its specific angular
   momentum and mass obey (\ref{qdef}) for any $p$.

3. For $\vec q=0$, there is a $p_*$ such that a black hole forms if
   and only if $p>p_*$. 

We should note that $p_*$ is the value of $p$ at the threshold only
for $\vec q= 0$. The black hole threshold is locally described by
$p=p_{\rm crit}(\vec q)$ with $p_{\rm crit}(0)=p_*$.

Do data sets from such a family go through the intermediate linear
regime (\ref{intermediate_linear})? 

a) If $\vec q=0$ corresponds to spherical symmetry, and if $|\vec q|$
is sufficiently small, the answer is yes. In this case, the deviations
from spherical symmetry are described by perturbation theory
throughout. We can then appeal to the experimental fact that in
spherical symmetry, the basin of attraction of the spherically
symmetric critical solution is the entire black hole threshold,
including data that are far from the critical solution. 

b) If $\vec q$ does not correspond to spherical symmetry, the answer
is unknown. For example, there is no guarantee that the spherically
symmetric critical solution is reached from near-critical data that
are octant-symmetric (and therefore non-rotating) but highly oblate or
prolate. But it is suggested by the fact that a spherical critical
solution exists, and that critical solutions apparently have as much
symmetry as possible. (In critical collapse in pure gravity, the
critical solution cannot be spherically symmetric, but it is
axisymmetric). 


\section{Universal scaling functions}
\label{section:scalingfunctions} 


We are now ready to calculate $\vec a(p,\vec q)$ and $M(p,\vec q)$.
The amplitudes of $Z_0$ and $\vec Z_1$ during the intermediate linear
regime depend on ${p}$ and ${\vec q}$ in some complicated, nonlinear,
and family-dependent way. Assuming that the dependence is analytic, we
can determine the behavior to leading order from symmetry
considerations. From the definition (\ref{qdef}), and because $\vec
Z_1$ carries the angular momentum during the intermediate linear
regime, $\vec B$ is an odd function of ${\vec q}$, and $A$ is an even
function. Also, both $A$ and $\vec B$ must vanish for ${p}=p_*$ and
${\vec q}=0$ because by definition that initial data set is on the
black hole threshold. To leading order in $p$ and $\vec q$, we find
that
\begin{equation}
\label{ansatz}
Z(x,\tau) \simeq Z_*(x) + C_0 ({{p}}-p_*)\, e^{\lambda_0\tau}Z_0(x) +
(C_1 {{\vec q}})\cdot e^{\lambda_1\tau}\vec Z_1(x,\Omega).
\end{equation}
Here $C_0$ and $C_1$ are two unknown constants that depend on the
two-parameter family we consider. $C_1$ is a $3\times 3$ matrix. In
axisymmetry it reduces to a number, but in general it is a non-trivial
map between the vector $\vec q$ and the vector $\vec Z_1$, and
therefore, in the end, between $\vec q$ and $\vec a$.

We now define $\tau_*$ such that
\begin{equation}
\label{epsilon}
C_0|{{p}}-p_*| e^{\lambda_0\tau_*}\equiv\epsilon, 
\end{equation}
where $\epsilon$ is a fixed small constant. We then have
\begin{equation}
\label{data}
Z(x,\tau_*) \simeq Z_*(x) \pm \epsilon Z_0(x) + \vec\delta\cdot
\vec Z_1(x,\Omega),
\end{equation}
where
\begin{equation}
\label{calcdelta}
\vec\delta \equiv C_1 {\vec q} \left(\epsilon\over C_0
|{p}-p_*|\right)^{\lambda_1\over\lambda_0} .
\end{equation}
Here the sign in front of $\epsilon$ is that of $p-p_*$. It appears
because of the absolute value taken in the definition
(\ref{epsilon}). Recall that in the following $\epsilon$ is a fixed
positive constant, while $\vec\delta$ depends on the initial data.

In order to simplify our notation, it is convenient to
introduce ``reduced'' values $\bar{p}$ and $\bar{\vec q}$ of ${p}$ and
${\vec q}$ as
\begin{equation}
\label{reduced}
\bar {p} \equiv C_0 \epsilon^{-1} (p-p_*),
\qquad
\bar {\vec q} \equiv C_1 \vec q,
\end{equation}
Note that by definition the direction of $\bar{\vec q}$ is now the
direction of $\vec a$. In the following, this is implied, and we
suppress the vector indices. In this simplified notation,
\begin{equation}
\label{definedelta}
\delta =
|\bar {p}|^{-{\lambda1\over\lambda_0}} \bar q, 
\end{equation}
The parameters $\bar {p}$ and $\bar q$ can be thought of as
``reduced'' values of ${p}$ and ${\vec q}$, which hide the
family-dependent part of the dependence of $M$ and $\vec a$ on $p$ and
$\vec q$. They are similar to the ``reduced temperature'' and
``reduced external field'' in statistical mechanics. We can also think
of them as coordinates on the phase space, in the way already
discussed for $\bar p$. 

We now consider $Z(x,\tau_*)$ as Cauchy data for a time evolution in
$\tau$ that leads away from the perturbative regime and eventually to
black hole formation or dispersion. $Z(x,\tau_*)$ contains the
complete Cauchy data, up to an overall scale, which must be provided
separately. That scale is $e^{-\tau_*}$, which from our definitions is
proportional to $|\bar {p}|^{-1/\lambda_0}$. In particular, the mass
of the black hole that is formed must be proportional to this scale,
with a constant of proportionality that can depend only on $\delta$
and the sign of $\bar p$. We find
\begin{equation}
\label{M}
M({{p}},{{q}}) \simeq |\bar {p}|^{1\over\lambda_0} \cases{ F_M^+
(\delta), & $\bar
{p} > 0$ \cr F_M^- (\delta), & $\bar {p} < 0$ .  }
\end{equation}
The functions $F_M^\pm(\delta)$ are universal, and so are the two
exponents $\lambda_0$ and $\lambda_1$. Because of (\ref{qdef}), $F_M$
is an even function.

Consider the special case $\vec q=0$, and hence $\delta=0$. Then only
the mode $Z_0$ is present in the intermediate linear regime, and, up
to scale, we only need to consider the two data sets $Z_*-\epsilon
Z_0$ and $Z_*+\epsilon Z_0$. It is known that the first data set
disperses, while the second one forms a black hole. If we use $M=0$ to
denote the absence of a black hole, we have $F_M^-(0)=0$ and
$F_M^+(0)>0$. We use the freedom to normalize $Z_0(x)$ in order to set
$F_M^+(0)=1$ as a convention. We then find
\begin{equation}
\label{Mpower}
M({{p}},{{q}}) \simeq \cases{ {\bar {p}}^{1\over\lambda_0}, & $\bar
{p} > 0$ \cr 0, & $\bar {p} < 0$ }\qquad \hbox{as $\bar {q}\to 0$.}
\end{equation}
We have recovered the power law in its form (\ref{simplepower2}), with
the critical exponent $\gamma=1/\lambda_0$ for the black hole mass.

We now consider the specific angular momentum $\vec a$ of the black hole. The
dimensionless quantity $a/M$ can depend on the initial data only
through the dimensionless quantity $\delta$, and on the sign of
$\bar{p}$. $M$ itself is given by (\ref{M}), and we only have to put
these two results together to obtain $a$. It is convenient to absorb
the scaling function $F_M(\delta)$ into the new scaling function for
$a$, and we can write
\begin{equation}
\label{a}
a({{p}},{{q}}) \simeq |\bar {p}|^{1\over\lambda_0} \cases{ 
F_a^+ (\delta),
& $\bar {p} > 0$ \cr
F_a^- (\delta),
& $\bar {p} < 0$ 
} 
\end{equation}
Here $F_a^\pm(\delta)$ are two new universal scaling
functions. Clearly, $F^\pm_a(0)=0$. Recall that we have adapted a
simplified notation, in which $F_a$ and $\delta$ are really vectors,
but in which $F_a$ takes its direction trivially from $\delta$, so
that $F_a$ is really just a scalar function of a scalar
variable. Because of (\ref{qdef}), it is an odd function. If cosmic
censorship holds, then $a<M$ must hold in black holes formed in
gravitational collapse, and therefore $|F_a|<F_M$ for all $\delta$.

To leading order $F_a$ is proportional to $\delta$.  By adjusting the
normalization of $\vec Z_1$, we can set the factor of proportionality
to one. With this convention, and from the definition
(\ref{definedelta}), we obtain a power law for $a$ in the limit of
small $\bar q$:
\begin{equation}
\label{apower}
a({{p}},{{q}}) \simeq 
\cases{ 
\bar {q}\ {\bar {p}}^{1-\lambda_1\over\lambda_0}, & $\bar {p} > 0$ \cr
0, & $\bar {p} < 0$  }\qquad \hbox{as $\bar {q}\to 0$.}
\end{equation}
This had been obtained previously under the assumption that
$\lambda_1<0$ \cite{angmom}.


\section{Proposed numerical tests} 
\label{section:tests}


So far our results have been formal: we know the two exponents
$\lambda_0$ and $\lambda_1$, but not the universal scaling
functions. However, we predict that such functions exist and are
universal. This can be tested by comparing two or more two-parameter
families of initial data. The two family-dependent parameters
$C_0/\epsilon$ and $C_1$ (the latter is a matrix) must be determined
for each family, by fitting observations in the limit ${q}\to 0$ to
the formulas (\ref{Mpower}) and (\ref{apower}). The scaling functions
can then be read off from the first family, and tested against the
second and any further family.

Alternatively, the scaling functions can be calculated directly by
evolving the data (\ref{data}) for fixed $\epsilon$ and all values of
$\delta$. In the notation of \cite{critfluid2}, the area radius $r$ in
spherical symmetry is related to the dimensionless radial coordinate
$x$ and the scale coordinate and coordinate $\tau$ by $r\equiv
sxe^{-\tau}$. ($s$ is a known constant whose significance does not
matter here.) We set $\tau=0$, and consider the initial data
\begin{equation}
\label{numdata}
Z_{\rm initial}(r,\Omega)=Z_*(r/s)\pm AZ_0(r/s)+BZ_1(r/s,\Omega).
\end{equation}
Here $Z_*(x)$ is already known, and $Z_0(x)$ and $Z_1(x)$ are known up
to normalization \cite{critfluid2}. From the three-fold degenerate
$\vec Z_1$, we arbitrarily choose $m=0$. The actual dimensionless
variables that $Z$ stands for in perfect fluid collapse are defined in
\cite{critfluid2}. The physical, dimensionful fluid variables of
\cite{fluidpert} are obtained from the rescaled variables defined in
\cite{critfluid2} by setting $\tau=0$ everywhere. 

We now evolve these data nonlinearly in axisymmetry. In order to
explore all values of $\delta$ in the universal scaling functions, we
could keep $A$ fixed at a small value $\pm\epsilon$, and vary $B$ from
$0$ to $\infty$. In practice, however, we want to vary the ratio $A/B$
from $-\infty$ to $\infty$, but choose an overall factor in both $A$
and $B$ small enough so that the initial data are in the intermediate
linear regime but large enough so that the evolution leaves this
regime soon. This minimizes the range of scales that the nonlinear
evolution code has to cover, and may make the calculation possible
without the need for adaptive mesh refinement.

Clearly, $\bar p$ is proportional to $A$ and $\bar q$ is proportional
to $B$. The constants of proportionality, however, are unknown because
we do not have the correct normalizations of $Z_0$ and $Z_1$. We just
fix an arbitrary normalization, and put in two adjustable
constants. Then the mass and specific angular momentum of the final
black hole that is created from the initial data (\ref{numdata}) are
related to the scaling functions by
\begin{eqnarray}
M(\pm \alpha A,\beta B)&=&A^{1\over\lambda_0}
F_M^\pm\left(B A^{-{\lambda_1\over\lambda_0} }\right), \\ 
a(\pm \alpha A,\beta B)&=&A^{1\over\lambda_0} F_a^\pm\left(B
A^{-{\lambda_1\over\lambda_0} }\right).
\end{eqnarray}
The constants $\alpha$ and $\beta$ correspond to the unknown correct
normalizations of $Z_0$ and $Z_1$, and must be adjusted to obtain the
two conventions
\begin{equation}
\label{conventions}
F_M^+(0)=1,\qquad {F_a^+}'(0)=1.
\end{equation}


\section{Large angular momentum} 
\label{section:largeangmom}


Current axisymmetric rotating fluid evolution codes are not ready at
the time of writing to calculate the universal scaling functions in
the manner suggested in the previous section. It is therefore tempting
to speculate about the form of the universal scaling functions at
large $\delta$.

Consider the limit where $\bar {p}\to 0$ while $\bar {q}\ne 0$ so that
$\delta\to \infty$. In physical terms this corresponds to first
fine-tuning a one-parameter family of initial data without angular
momentum to the black hole threshold and then adding a small but
finite amount of angular momentum to the data. Do the data obtained in
this way form a black hole? If our model is correct, the answer is
universally yes or no, independently of the family of zero angular
momentum data we have fine-tuned and independently of how we have
added that bit of angular momentum. The reason is that such data are
funneled through the data set $Z_*+\varepsilon \vec Z_1$ (for some
small fixed $\varepsilon$) in the intermediate linear regime. We now
discuss the two possible final states for these data in turn.

{\bf Possibility 1:} The data $Z_* + \varepsilon \vec Z_1$ form a
black hole. As its mass is finite (these data possess an intrinsic
scale), the explicit power of $\bar {p}$ in formula (\ref{M}) must be
canceled by a power-law behavior of $F_M(\delta)$ as
$\delta\to\infty$. One easily sees that this power law must be
\begin{equation} 
\label{alt1_FM}
F_M^\pm(\delta) \simeq C_M |\delta|^{1\over\lambda_1}, \quad
|\delta|\to\infty
\end{equation}
for the cancellation to occur. But then one immediately obtains that
\begin{equation}
M\simeq C_M |\bar {q}|^{1\over\lambda_1}, \qquad 
|\delta|\to\infty.
\end{equation}
By the same argument we have
\begin{equation} 
\label{alt1_Fa}
F_a^\pm(\delta) \simeq C_a \, {\rm sign}(\delta)\, |\delta|^{1\over\lambda_1},
\quad |\delta|\to\infty,
\end{equation}
and therefore
\begin{equation}
\label{critsus}
a \simeq C_a \, {\rm sign}(\bar {q})\, |\bar
{q}|^{1\over\lambda_1},\qquad 
|\delta|\to\infty.
\end{equation}
The constants $C_a$ and $C_M$ are again universal. We see that $a/M$
goes to the constant $C_a/C_M$ as we increase the angular momentum in
the initial data. We do not know what the value of this constant is,
except that cosmic censorship requires $C_a/C_M\le 1$.

{\bf Possibility 2:} The data $Z_* + \varepsilon \vec Z_1$ do not form a black
hole. On physical grounds this alternative appears more likely, as one
would expect centrifugal forces to disrupt data that already hover
between collapse and dispersion. This means that $F_{M,a}^\pm(\pm
\infty) = 0$. If we consider obtaining this limit by adding angular
momentum to supercritical data, that is by increasing $\bar {q}$ at
fixed $\bar {p}>0$, it appears likely that above a threshold amount of
angular momentum no black hole is formed at all, rather than ever
smaller ones. In this case it also seems plausible that zero angular
momentum data below the black hole threshold will not form a black
hole if one adds angular momentum. If these two assumptions are true,
then instead of (\ref{alt1_FM}) and (\ref{alt1_Fa}) we have 
\begin{equation}
F_{M,a}^+(\delta) = 0 \quad \hbox{for $|\delta|>\delta_{\rm
max}$}, \quad F_{M,a}^-(\delta)=0 \quad \hbox{for all $\delta$}.
\end{equation}
This means that black holes are formed if and only if
\begin{equation}
\label{critsurface}
 \bar {p}>0 \quad \hbox{and} \quad |\bar {q}|
< \delta_{\rm max} \ {\bar {p}}^{\lambda1\over\lambda_0}.
\end{equation}
This defines a convex region in the ${p},{q}$ plane. 
We do not know if black holes at the threshold with non-zero angular
momentum are formed with zero or finite mass. This gives rise to two
sub-cases:

{\bf Possibility 2a:} In the presence of angular momentum there is a
mass gap at the black hole threshold. This means that $F_M^+$ is
discontinuous at $\delta_{\rm max}$, of the form
\begin{equation}
F_M^+(\delta)\simeq \cases{K_M, & $\delta\lesssim \delta_{\rm max}$ \cr
0, & $\delta>\delta_{\rm max}$ \cr}
\end{equation}
for some universal constant $K_M$
and the size
of the mass gap is
\begin{equation}
\Delta M \simeq K_M\, {\bar {p}}^{1\over\lambda_0} 
\end{equation}
Similar behavior would hold for $F_a^+$, for some universal constant
$K_a$ so that
\begin{equation}
\Delta a \simeq K_a\, {\bar {p}}^{1\over\lambda_0} 
\end{equation}
Therefore, the spin/mass ratio of black holes formed at the threshold
is universal, with
\begin{equation}
{\Delta a\over \Delta M}= {K_a\over K_M}.
\end{equation}
From cosmic censorship we have $K_a/K_M<1$. Fig. \ref{figure:case2a}
gives a qualitative impression of $M(\bar p,\bar q)$ and $a(\bar
p,\bar q)$ in case 2a.

{\bf Possibility 2b:} There is no mass gap even in the presence of
angular momentum. This means that $F_M^+(\delta)$ vanishes at
$\delta_{\rm max}$. If this happens, for example, as a power $n$ of
distance from the threshold,
\begin{equation}
F_M^+(\delta)\simeq \cases{K_M (\delta_{\rm
max}-\delta)^n , & $\delta\lesssim \delta_{\rm max}$ \cr
0, & $\delta>\delta_{\rm max}$ \cr}
\end{equation}
then the black hole mass would scale as that same power $n$ of any
regular coordinate scalar on the phase space  that vanishes at the
threshold (except at $\vec q=0$). Fig. \ref{figure:case2b}
gives a qualitative impression of $M(\bar p,\bar q)$ and $a(\bar
p,\bar q)$ in case 2b.


\section{The analogy with statistical mechanics} 
\label{section:statmechanalogy}


This work was motivated by the attempt to exploit the critical phase
transition/critical collapse analogy to learn something new about
critical collapse. The preceding material has been presented in a
self-contained manner without any explicit reference to statistical
physics concepts, but it may be interesting to point out the exact
parallels now.

The calculation of critical exponents in critical collapse is
mathematically identical to the calculation of critical exponents in
statistical mechanics, even though the underlying physical phenomena
are totally different. The mathematical equivalent of the
renormalization group acting on micro-states in statistical mechanics
is the time evolution, in certain preferred coordinates, acting on
initial data in general relativity. Both can be considered as
dynamical systems. Critical exponents are calculated by linearizing
around their critical fixed points \cite{critreview2,Goldenfeld}.

In statistical mechanics, the critical fixed point typically has two
growing modes. Physically these are linked to the temperature $T$ and
to a generalized external force. It is helpful to consider two
contrasting examples.  For a fluid confined in a vessel so that a
vapor phase is in equilibrium with a liquid phase, the generalized
force is the pressure $P$, and the order parameter is the difference
$\rho_{\rm liquid}-\rho_{\rm gas}$ between the densities of the two
phases. For a ferromagnet, the generalized force is the external
magnetic field $\vec H$, and the order parameter is the magnetization
$\vec m$.  In both cases the temperature-force plane contains a line
of first-order phase transitions ending in a second-order phase
transitions. (In the standard terminology, a first-order transition is
one where the order parameter jumps from zero to a non-zero value. A
second-order, or critical, phase transition is one where the order
parameter is continuous, rising as a power-law.) For the fluid, this
is the liquid-gas phase transition, ending at the critical point where
these two phases have the same density. For the ferromagnet, the phase
transition is between the possible directions of the magnetization
$\vec m$ at $T<T_C$. This first-order transition ends at the Curie
temperature $T_C$. The spontaneous magnetization at zero external
field behaves as a power of $T_C-T$ for $T\lesssim T_C$.

If one restricts consideration to a vanishing external magnetic field,
one appears to have an analog of the black-hole threshold in spherical
symmetry, where the $T_C-T$ corresponds to $p-p_*$ and the absolute
value of the magnetization $|\vec m|$ corresponds to the black hole
mass $M$. The formula (\ref{simplepower}) corresponds to the
dependence of the spontaneous magnetization on the temperature, in the
absence of an external field.  But typical critical phase transitions
in statistical mechanics have two independent critical exponents,
corresponding to two growing modes of the fixed point. Nigel
Goldenfeld has asked if a second growing mode could not be found in
the critical collapse problem so that the universal critical behavior
would show not one critical exponent but two and, more interestingly,
universal scaling functions \cite{Goldenfeldcomm}.

Here we suggest that the analog of the external magnetic field $\vec
H$ in the collapse problem is the angular momentum parameter
$\vec{q}$. Angular momentum provides a second growing mode of the
critical point. The analog of the order parameter $\vec m$, a vector,
is not the black hole mass but its (specific) angular momentum vector
$\vec a$. Conversely, the analog of the black hole mass $M$ is a
scalar that scales like a length, for example the correlation length
$\xi$. (The fact that $\xi$ diverges at the critical point, while $M$
goes to zero because that is what length scales do in the two kinds of
critical point.) We have shown that a second critical exponent and
universal scaling functions do indeed arise, in exact mathematical
analogy with the ferromagnet model \cite{Goldenfeld}.

In both the ferromagnet and collapse models the two growing modes have
different symmetries. While $Z_*+\epsilon Z_0$ forms a black hole,
$Z_*-\epsilon Z_0$ disperses. The two signs are qualitatively
different. But $Z_*+\epsilon Z_0+\vec\delta \cdot\vec Z_1$ will either
collapse or disperse depending only on the absolute value of $\delta$,
independently of its direction. Similarly in the ferromagnet, the
state at $T=T_C+\epsilon$ differs qualitatively from that at
$T=T_C-\epsilon$, but only the absolute value of $\vec H$ matters.  In
each case, the vector-valued parameter ($\vec q$ or $\vec H$) is
associated with a symmetry breaking. Because $\vec q=0$ or $\vec H=0$
correspond to an unbroken symmetry, their critical value is obviously
zero, whereas the critical values $T_C$ and $p_*$ are
nontrivial. ($T_C$ depends on the material, and $p_*$ on the family.)
If one imposes the symmetry on the problem (no external magnetic
field, or octant symmetry in collapse), the critical point has
effectively only one unstable mode.

There is one major difference between the ferromagnet and critical
collapse: In the ferromagnet, an infinitesimal external field at a
temperature slightly above the Curie temperature creates a finite
magnetization. The ferromagnetic region is therefore concave in the
$T$-$H$ plane. Even with zero external field, the symmetry is
spontaneously broken, and a finite net magnetization in a random
direction results. Fig. {\ref{figure:ferromagnet} gives a qualitative
impression of $m(T_C-T,H)$ for a ferromagnet in axisymmetry, and
should be contrasted with Figs. \ref{figure:case2a} and
\ref{figure:case2b}. A partial analog of this in critical collapse
would be possibility 1 in the previous section. However, possibility 2
appears more likely on the physical grounds that centrifugal forces
should oppose collapse. The black hole region is then convex in the
$\bar{p}\bar{q}$ plane.

The analogy between critical collapse and the critical point of a
fluid is less close than between critical collapse and the
ferromagnet, as neither temperature nor pressure are associated with a
symmetry breaking: both have to be fine-tuned at once to nontrivial
critical values that depend on the material. As a curiosity, however,
we note that an analog of the liquid-gas transition, where $q$ is a
scalar parameter, and not associated with a symmetry breaking, could
arise in critical collapse in the context of semiclassical
gravity. Self-similarity in critical collapse in semiclassical gravity
is broken at small scales (of the order of the Planck scale) by the
quantum stress-energy. Brady and Ottewill \cite{BO} suggest that the
effect of the quantum stress-energy can be described approximately in
terms of a single unstable mode (in addition to the known classical
one) with a growth exponent of $\lambda_1=2$. If this description is
correct, the parameter $q$ would describe some measure of the
excitation state of the quantum fields.


\section{Conclusions} 
\label{section:conclusions}


The black hole threshold in gravitational collapse has been
investigated until now only for initial data without angular
momentum. It is largely unknown what happens when the threshold is
crossed along a family of initial data with nonzero angular
momentum. If the angular momentum is very small, it can be treated as
a perturbation of spherical collapse throughout, and for this case a
critical exponent for the black hole angular momentum has been
predicted \cite{angmom,critfluid2,critscalangmom} and is now waiting
to be tested in numerical collapse simulations.

Here we have noted that the perturbative calculations giving rise to
this prediction can be extended in the case where the angular momentum
perturbation constitutes a second growing mode of the self-similar,
spherically symmetric critical solution, in addition to the spherical
growing mode known already. We can now extend our predictions to a
regime where the angular momentum of the initial data is small, but
large enough so that the final black hole is rapidly rotating.

In Eqs. (\ref{M}) and (\ref{a}) we have made predictions for the black
hole mass and specific angular momentum in this regime, expressed in terms of
two known critical exponents and two yet unknown functions of one
variable $\delta$. The prediction that universal scaling functions
exist at all can be tested by comparing two or more two-parameter
families of initial data. The scaling functions could also be
calculated more simply and precisely through the nonlinear evolution
of only two one-parameter families of axisymmetric initial.

The necessary initial data are already known to high precision, and
the necessary technology (an axisymmetric fluid code that can form
rotating black holes, but without mesh refinement) should be available
soon. In the meantime we have discussed the qualitative features of
these functions, distinguishing two possibilities, and suggesting one
of them as the most plausible on physical grounds.

The prediction of universal scaling functions in critical collapse
with angular momentum both extends and clarifies the analogy between
critical phenomena in gravitational collapse and critical phase
transitions in statistical mechanics, in particular the ferromagnetic
phase transition.


\acknowledgments

I am grateful to Nigel Goldenfeld and Bob Wald for interesting
discussions, and to Jos\'e M. Mart\'\i n-Garc\'\i a and James Vickers for
comments on the manuscript. This research was funded by in part by NSF
grant PHY-9514726 to the University of Chicago, and by EPSRC grant
GR/N10172.



\break 

\begin{figure}
\epsfxsize=12cm
\centerline{\epsffile{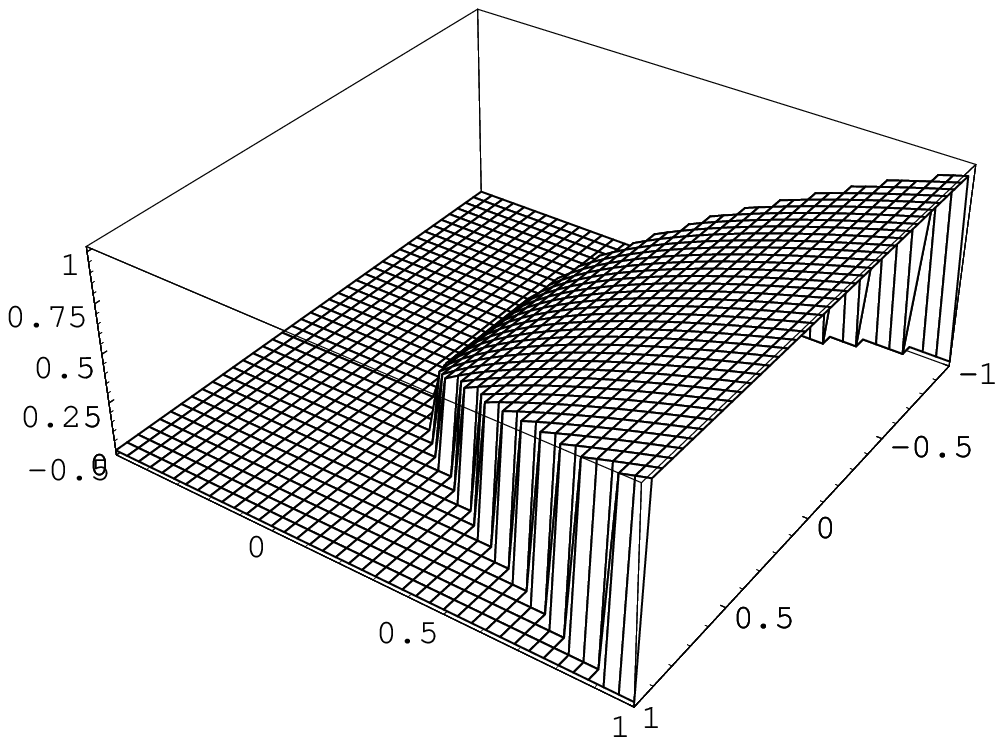}}
\epsfxsize=12cm
\centerline{\epsffile{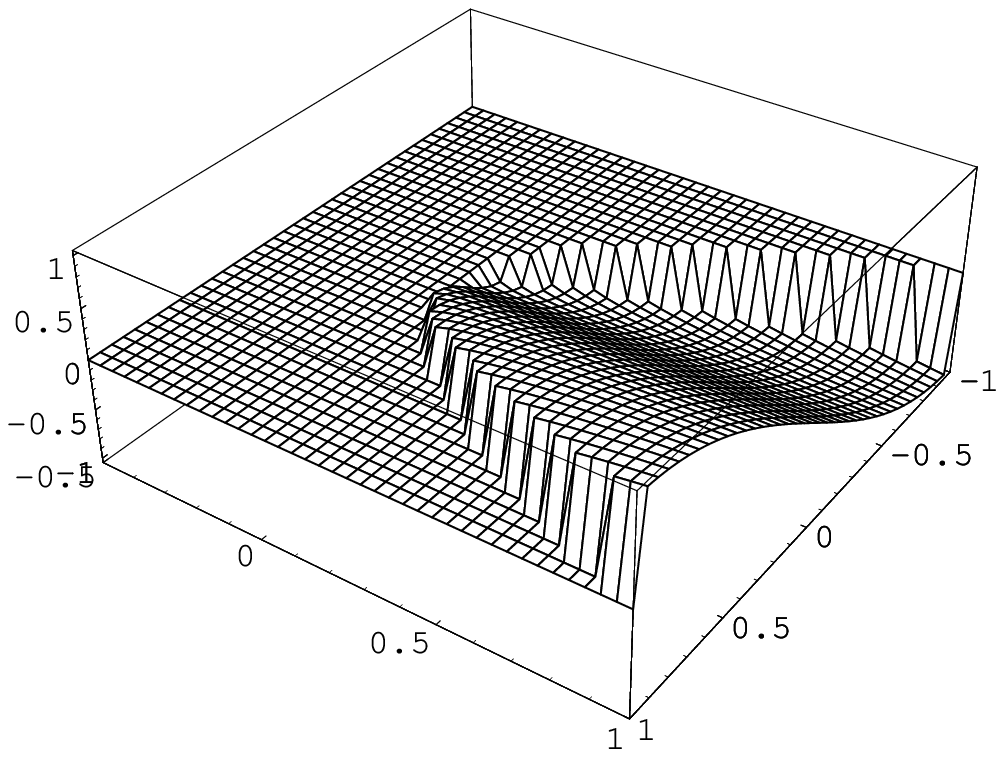}}
\caption{\label{figure:case2a} Schematic plot of black hole mass $M$
and specific angular momentum $a$ in axisymmetry as functions of $\bar
p$ and $\bar q$, assuming case 2a. For illustration we have assumed
$\lambda_0=2$, $\lambda_1=1$, $F^+_M(\delta)=\theta(1-|\delta|)$ and
$F^+_a(\delta)=\theta(1-|\delta|)\sin(\pi\delta/2)$.  Note that in
reality $\lambda_1/\lambda_0\ll1$, and that we have assumed that
$|a|/M^2\to 1$ at the black hole threshold.}
\end{figure}

\break 

\begin{figure}
\epsfxsize=12cm
\centerline{\epsffile{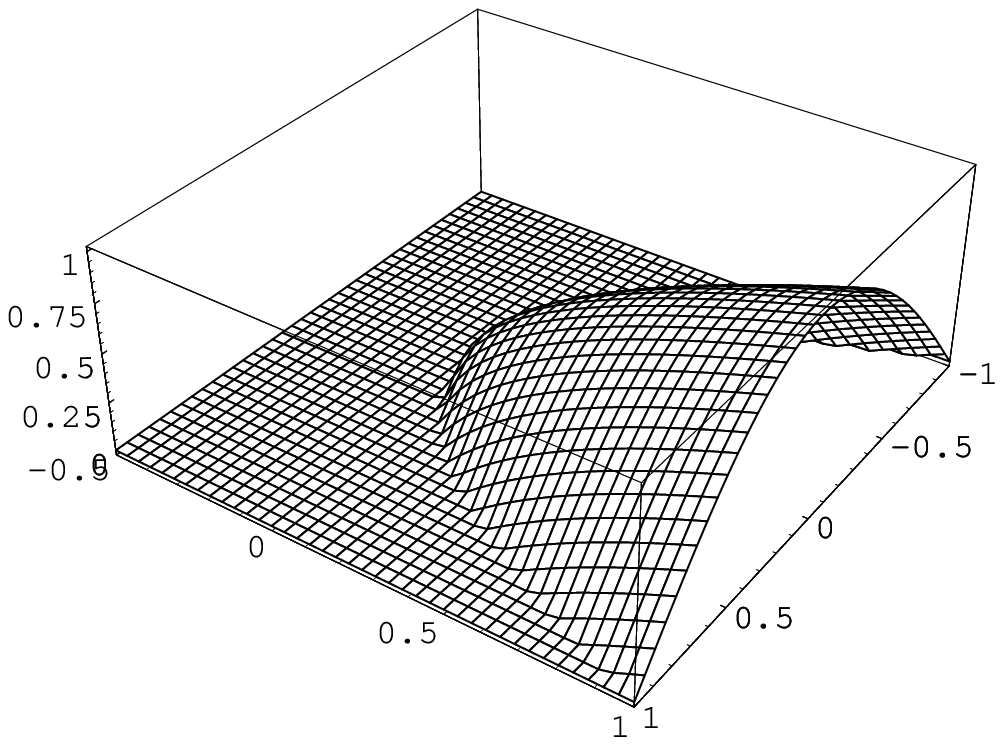}}
\epsfxsize=12cm
\centerline{\epsffile{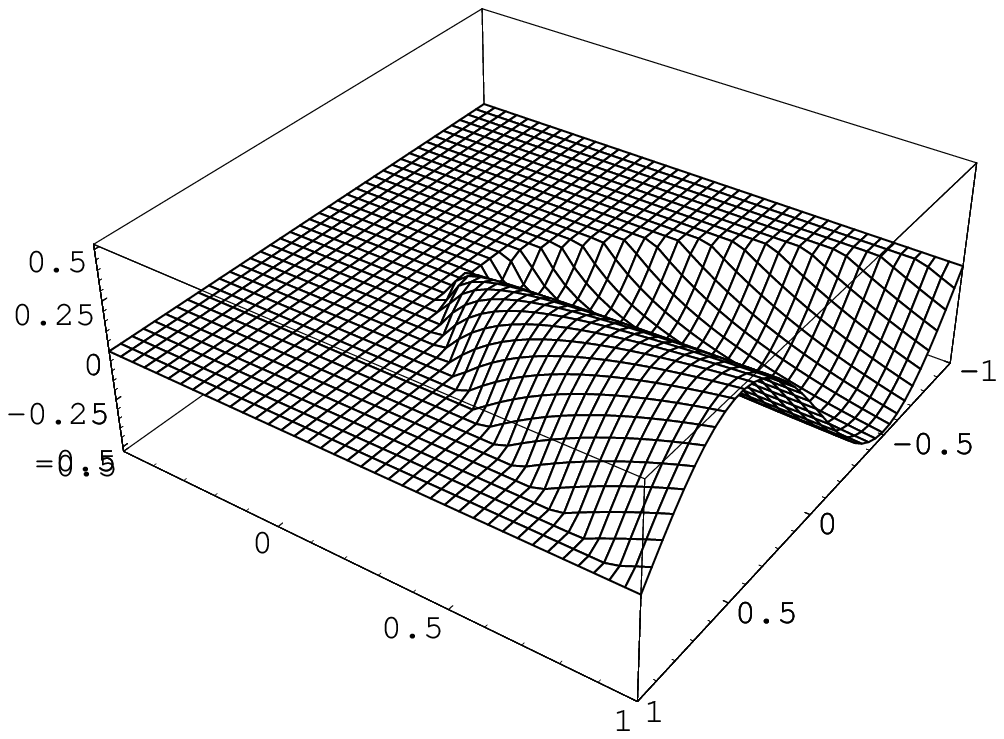}}
\caption{\label{figure:case2b} Schematic plot of black hole mass $M$
and specific angular momentum $a$ in axisymmetry as functions of $\bar
p$ and $\bar q$, assuming case 2b. For illustration we have assumed
$\lambda_0=2$, $\lambda_1=1$,
$F^+_M(\delta)=\theta(1-|\delta|)\cos(\pi\delta/2)$ and
$F^+_a(\delta)=\theta(1-|\delta|) \sin(\pi\delta) /2$. Note that in
reality $\lambda_1/\lambda_0\ll1$, and that we have assumed that
$|a|/M^2\to 1$ at the black hole threshold.}
\end{figure}

\break

\begin{figure}
\epsfxsize=12cm
\centerline{\epsffile{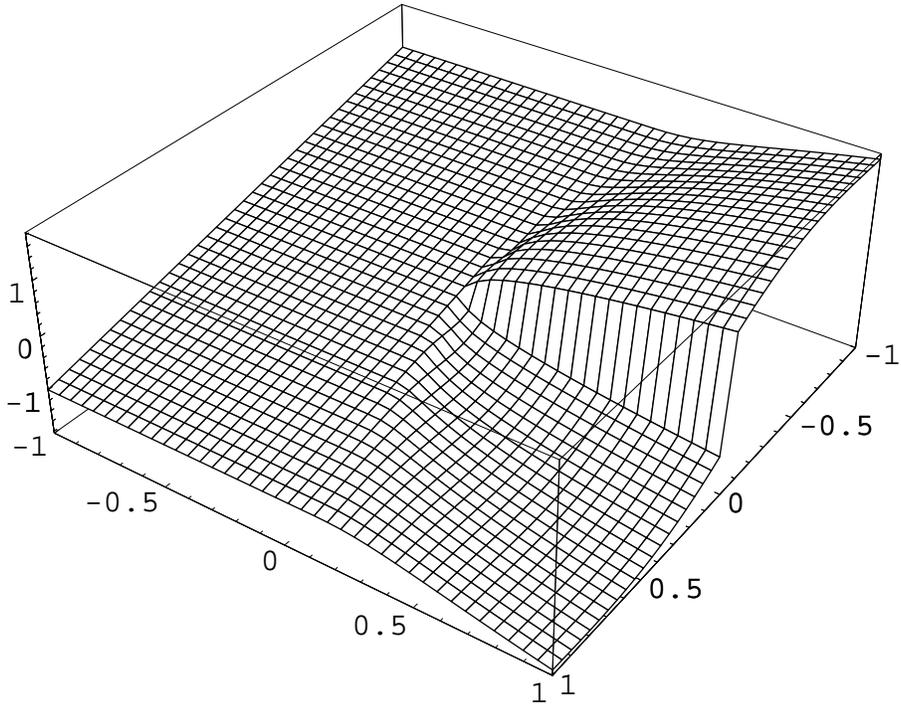}}
\caption{\label{figure:ferromagnet} Schematic plot of the
magnetization $m$ of a ferromagnet as a function of $T_C-T$ and
$H$. For illustration we have assumed $\lambda_0=2$, $\lambda_1=1$,
$F^-(\delta)=\rm{sign}(\delta)|\delta|^{1/\lambda_1} $ and
$F^+(\delta)=F^-(\delta)+e^{-\delta^2}[\rm{sign}(\delta)+\delta]$.  }
\end{figure}


\end{document}